\documentclass[12pt]{article}
\usepackage{amsmath}
\usepackage{amssymb}
\usepackage{amsthm}
\usepackage{mathtools}
\usepackage{url}
\usepackage[usenames,svgnames]{xcolor}
\usepackage{enumitem}
\usepackage{comment}
\usepackage{mathdots}
\usepackage{graphicx}
\usepackage{float}
\usepackage{enumitem}
\usepackage[symbol]{footmisc}
\usepackage[pagebackref=false]{hyperref} %hyperref wants to be loaded last. Set = false if no pagebackref needed
\usepackage[all]{hypcap} %fix hyperref links to tables/figures/etc.

\theoremstyle{plain}
\theoremstyle{definition}
\newtheorem{theorem}{Theorem}[section]

\newtheorem{definition-theorem}[theorem]{Definition-Theorem}
\newtheorem{definition-proposition}[theorem]{Definition-Proposition}

\newtheorem{example}{Example}[section]
\newtheorem{examples}{Example}[subsection]

\newtheorem{remark}{Remark}[section]

\newtheorem{definition}{Definition}[section]

\numberwithin{equation}{section} % requires package amsthm

\DeclareMathOperator{\Span}{span}

\DeclareMathOperator{\diag}{diag}

\def \pa{\partial}

\def\tr{\mathrm {tr}}
\def\det{\mathrm {det}}

\def\Ad{\mathrm {Ad}}

\def\diag{\mathrm {diag}}

\def\res{\mathop{\mathrm {res}}\limits}

\def\res{\mathop{\mathrm{res}}\limits}

\def\&{&{\hskip -20pt}}

\def\be{\begin{equation}}
\def\ee{\end{equation}}

\def\bea{\begin{eqnarray}}
\def\eea{\end{eqnarray}}

\def\bt{\begin{theorem}}
\def\et{\end{theorem}}

\def\bex{\begin{example}\small \rm}
\def\eex{\end{example}}

\def\bexs{\begin{examples}\small \rm}
\def\eexs{\end{examples}}

\def \ss {\subset}

\def\br{\begin{remark}\small \rm \em}
\def\er{\end{remark}}

\def\DD {{\mathcal D}}

\def\II {{\mathcal I}}

\def\LL{{\mathcal L}}

\def\OO{{\mathcal O}}

\def\TT {{\mathcal T}}

\def\WW{{\mathcal W}}

\def\Cb{{\mathbf C}}

\def\Pb{{\mathbf P}}

\def\Tb{{\mathbf T}}

\def\Xb{{\mathbf X}}

\def\db{{\mathbf d}}

\def\Ib{{\mathbf I}}

\def\Gl{\mathbf {Gl}}

\def\0b{\boldsymbol{0}}

\def\Cbb{\mathbb{C}}

\def\grGl{\mathfrak{Gl}} \def\grgl{\mathfrak{gl}}
 
 \def\grsl{\mathfrak{sl}}

\begin{document}
\baselineskip 16pt

\medskip
\begin{center}
\begin{Large}\fontfamily{cmss}
\fontsize{17pt}{27pt}
\selectfont
	\textbf{Hamiltonian structure of isomonodromic deformation dynamics in linear systems of PDE's}
	\end{Large}
	
\bigskip \bigskip
\begin{large}  
J. Harnad$^{1, 2}$\footnote[2]{e-mail:harnad@crm.umontreal.ca}  
 \end{large}
 \\
\bigskip
\begin{small}
$^{1}${\em Centre de recherches math\'ematiques, Universit\'e de Montr\'eal, \\C.~P.~6128, succ. centre ville, Montr\'eal, QC H3C 3J7  Canada}\\
$^{2}${\em Department of Mathematics and Statistics, Concordia University\\ 1455 de Maisonneuve Blvd.~W.~Montreal, QC H3G 1M8  Canada}\\
\end{small}
 \end{center}
\medskip

%%%%%%%%%%%%%%%%  Abstract  %%%%%%%%%%%%%%%%
\begin{abstract}
\footnotesize{
The Hamiltonian approach to isomonodromic deformation systems for generic rational covariant derivative operators on the Riemann sphere having
any matrix dimension $r$  and any number of isolated singularities of arbitrary Poincar\'e rank is derived using the split classical rational $R$-matrix
Poisson bracket structure on the dual space $L^*\grgl(r)$ of the loop algebra $L\grgl(r)$. Nonautonomous isomonodromic counterparts of isospectral systems  are obtained by identifying the deformation parameters as Casimir elements on the phase space. These are shown to coincide with the higher Birkhoff invariants determining the local asymptotics near to irregular singular points, together with the pole loci. They appear as the negative power coefficients in the principal part of the Laurent expansion of the fundamental meromorphic differential on the associated spectral curve, while the corresponding dual spectral invariant Hamiltonians appear as the ``mirror image'' positive power terms. Infinitesimal isomonodromic deformations are  generated by the sum of the Hamiltonian vector field and an {\it explicit derivative} vector field that is transversal to the symplectic foliation. }
 \end{abstract}

%%%%%%%%%% Section 1. Introduction: Isomonodromic deformations %%%%%%
\normalsize
 \section{Introduction: Isomonodromic deformations}

 \subsection{Isomonodromic deformation systems history}

1889-1912:  Isomonodromic deformations of  linear differential systems having a finite number of isolated singular
   points with finite pole degrees were studied since the pioneering work of  Picard \cite{Pic},  Painlev\'e \cite{Pa1, Pa2}, Fuchs \cite{Fu},  Gambier \cite{Gam}, 
   Schlesinger \cite{Sch}, Garnier \cite{Gar} and others.
   
   \medskip\noindent
1913: An extension of the concept of monodromy to include systems with irregular isolated singularities  and generalized monodromy data  ({\it Stokes} and {\it connection} matrices) was developed by Birkhoff \cite{Birk}. 

\medskip\noindent   
1980-81: A revival of interest in isomonodromic deformations was  inspired by advances in the theory of  completely integrable systems, 
which led to new results by Flaschka and Newell \cite{FN1, FN2} and Jimbo, Miwa and Ueno \cite{JMU, JM}, 
who introduced the notion of {\it isomonodromic $\tau$-functions}.

\medskip  
 The presence of a {\it Hamiltonian structure} underlying such deformation equations was recognized since early studies of the
 Painlev\'e equations (Fuchs (1905) \cite{Fu}, Painlev\'e (1906) \cite{Pa2}, Malmquist (1922) \cite{Malm}, Okamoto (1980) \cite{Ok}),
  and related to the notion of isomonodromic $\tau$-functions by Jimbo, Miwa, Ueno (1981) \cite{JMU, JM}. This was extended  to more 
  general rational systems using the classical rational $R$-matrix Poisson bracket structure on loop algebras 
  (Harnad (1994) \cite{H1},  Boalch (2001) \cite{Bo1, Bo2}). For an introduction to the Hamiltonian structure of  rational isomonodromic 
  deformations equations, the corresponding $\tau$-functions and applications, see \cite{HB}, Chapts. 9-12. Detailed accounts of various approaches to the 
  Hamiltonian structure of isomonodromic deformations of covariant derivative operators on the punctured Riemann sphere with rational coefficients 
   may be found in refs. \cite{Yam, BHH, MOA, MA}.
   
  The following is an overview and summary of results on the Hamiltonian structure of isomonodromic deformation dynamics
for covariant derivative operators  of arbitrary rank with rational coefficients on the Riemann sphere, based largely on \cite{H1, H2, BHH}.

\subsection{Rational linear differential systems}
  
Consider covariant derivative operators
\be
\DD_z^L := \frac{\pa}{\pa z} - L(z),   \quad z\in \Cb,
\label{rational_covar_deriv}
\ee
on the punctured Riemann sphere, where $L(z)$ is a rational $r \times r$ Lax matrix of the form
\bea
L(z)&\& =
 -\sum_{j=0}^{d_\infty-1} L^{\infty}_{j+2} z^j + \sum_{\nu=1}^N
\sum_{j=1}^{d_\nu+1} \frac {L^{\nu}_{j}}{(z-c_\nu)^j}  ,
  \\ 
&\& L_{d_\infty+1}^{\infty}\in \mathfrak h_{reg} \ss \grgl(r),  \  L_{d_\nu+1}^{\nu}\in {\frak g}_{reg} \ss \grgl(r), \ \ c_\nu\neq c_\mu, \ \nu\neq \mu,
\label{Ldef}
 \\ &\&
 \cr 
  r&\&= \text{rank}, \quad d_\nu= \text{Poincar\'e index}, \quad \db:=\{d_1, \dots, d_N, d_\infty\},
  \nonumber
\eea
Here $\mathfrak h_{reg}$ denotes diagonal $r \times r$ matrices with distinct eigenvalues and $ {\frak g}_{reg}$ denotes $\grgl(r)$ 
elements with distinct eigenvalues. The set of all such Lax matrices is denoted $\LL_{r, \db}$.
Let $\Psi(z) \in \Gl(r,\Cb)$ be a fundamental system of solutions to the linear system of first order ODE's
\be
 \frac{\pa \Psi(z)}{\pa z} = L(z) \Psi(z), \quad \Psi(z) \in \Gl(r,\Cb).
 \ee
 
 The results stated in the following two subsections  form the starting point of our analysis
 and are derived in detail in refs. \cite{Birk, BJL, JMU}.
     
  \subsection{Birkhoff deformation parameters}

\begin{theorem}[{\it Formal asymptotics and Birkhoff invariants}  \cite{Birk, BJL, JMU} ]
In terms of the local parameters 
\be
\zeta_\nu := (z-c_\nu), \quad \nu=1,\dots, N, \quad
\zeta_\infty := \frac 1 z , 
\ee
there exist local formal series solutions of the form 
\be
\Psi^\nu_{\text{\tiny form}}(z) =  
Y^\nu (\zeta_\nu) {\rm e}^{T^\nu(\zeta_\nu)} , \quad \ Y^\nu(\zeta_\nu) 
:= G^\nu\left(\Ib + \sum_{j\geq 1} Y_j^\nu {\zeta_\nu}^j\right),
\ee
in a punctured neighbourhood of each of the singular points $\{z=c_\nu\}$ , 
 where  $T^\nu(\zeta_\nu)\in \mathfrak h_{reg}$ is a  diagonal $r\times r$ matrix of the form 
\be
T^\nu (\zeta_\nu) = \sum_{j=1}^{d_\nu }\frac { T_j^\nu}{j {\zeta_\nu}^j} + T_0^\nu \ln \zeta_\nu, \quad  \ T^\nu_{d_\nu } = -(G^\nu)^{-1}L^\nu_{d_\nu+1} G^\nu,
\ee
for $\nu=1,\dots, N,\infty$. 
The columns of the invertible matrices $G^\nu\in GL(r,\Cb)$ 
  are the independent  eigenvectors of $L^\nu_{d_\nu+1}$ and $G^\infty=\Ib$.
  \end{theorem}

The notation used here for the diagonal values (the {\it Birkhoff invariants}) is:
\be
T^{\nu}_{j} = {\rm diag}(t^{\nu}_{j 1},\dots,t^{\nu}_{j  r}), \quad j=0, \dots, d_\nu 
\ee
so
\be
T^{\nu}(\zeta_\nu) = \sum_{j=1}^{d_\nu} \sum_{a=1}^r t^{\nu}_{j a} E_a \frac1{j\zeta_\nu^{j}} + \sum_{a=1}^r t^{\nu}_{0a}E_a \ln \zeta_\nu,
\ee
where
\be
t^{\nu}_{j a} \neq t^{\nu}_{j b} \ \  \text{for } a\neq b, \ \ j=1, \dots, d_\nu, \ \ \nu =1, \dots, N, \infty
\ee 
 and $E_a$ is the elementary $r \times r$ matrix whose only nonzero entry is a $1$ in the $(aa)$ position.
 \subsection{Jimbo-Miwa-Ueno isomonodromic deformation equations}
 
\begin{definition}[\cite{JMU}] The infinitesimal isomonodromic deformation matrices are defined as
\bea
U^\nu_{ja}(z; L)&\&:= \left(Y^\nu(\zeta_\nu) \frac {\pa  T^\nu(\zeta_\nu)}{\pa t^\nu_{ja}} (Y^\nu(\zeta_\nu))^{-1}\right)_{sing} \cr
&\& = \left(Y^\nu(\zeta_\nu) \frac { E_{aa} }{j \zeta_\nu^j } (Y^\nu(\zeta_\nu))^{-1}\right)_{sing},
\cr
V^\nu(z;L)&\& :=  \left( Y^\nu(\zeta_\nu) \frac {\pa  T^\nu(\zeta_\nu)}{\pa c_\nu } (Y^\nu(\zeta_\nu))^{-1}\right)_{sing} \cr
&\&=- \left( Y^\nu(\zeta_\nu) \frac {d T^{\nu}(\zeta_\nu)}{d z} (Y^\nu(\zeta_\nu))^{-1}\right)_{sing}
= -\sum_{j=1}^{d_\nu+1} \frac {L^\nu_{j}} {(z-c_\nu)^j}.
\eea
where  $(\cdot )_{sing}$ denotes the principal part of the Laurent series at a particular point $c_\nu\in \Pb^1$.
\end{definition}

\begin{theorem}[{\it Jimbo-Miwa-Ueno isomonodromic deformation equations} \cite{JMU}] 
If the following system of equations is satisfied 
\bea
\frac {d \Psi(z)}{d z}  &\& = L(z) \Psi(z),
\cr
\ \frac {\pa  \Psi(z)}{\pa t^\nu_{ja}} &\&= U^\nu_{ja}(z) \Psi(z) , \quad j=1, \dots, d_\nu ,
\cr
\frac {\pa  \Psi(z) }{\pa c_\nu}&\&= V^\nu(z) \Psi(z), \quad \nu =1, \dots , N,
 \label{JMUeq}
\eea
the generalized monodromy (including the values of the Stokes matrices defined in a neighbourhood of each irregular singular point) is independent of
the deformation parameters $\{t^\nu_{ja}, c_\nu\}$. 
\end{theorem}

\begin{remark}The exponents of formal monodromy $\{t^\nu_{0a}\}$ , \ $\nu=1, \dots , N, \infty$, \ do {\bf not} occur as deformation parameters.
\end{remark}

\begin{theorem}[{\it Consistency conditions: zero curvature equations} \cite{JMU}]
This overdetermined system is consistent if  the {zero curvature} equations are satisfied:
\bea
\frac{\partial L(z)}{\partial t^{\nu}_{ja}}&\&= \Big[U^\nu_{ja}(z), L(z)\Big] +\frac {dU^\nu_{j  a }(z) }{d z}, \cr
\frac{\partial L(z)}{\partial c_\nu}
&\&= \Big[V^\nu(z), L(z)\Big] +\frac {dV^\nu(z)}{d z}, \cr
&\& \cr
\frac{\partial  U^\mu_{k  b }(z) }{\partial t^{\nu}_{ja}} &\&= \Big[U^\nu_{ja}(z), U^\mu_{k  b }(z) \Big] 
+ \frac{\partial U^\nu_{j  a }(z)}{\partial t^{\mu}_{kb}} ,
\cr
\frac{\partial V^\mu(z)}{\partial c_\nu}
&\&
= \Big[V^\nu(z), V^\nu(z)\Big] +\frac {\partial V^\nu(z)}{\partial c_\nu},\cr 
\frac{\partial U^\mu_{k  b }(z) }{\partial c_\nu}  &\& =\Big[V^\nu(z), U^\mu_{k  b }(z) \Big] + \frac{\partial V^\nu(z)}{\partial t^{\mu}_{ja}}, \cr
\frac{\partial V^\mu(z)}{\partial t^{\nu}_{ja}} &\&= \Big[V^\nu(z), V^\mu(z)\Big] +  \frac{\partial V^\nu(z)}{\partial t^{\mu}_{ja}}. 
\eea
\end{theorem}

 %%%%%%%%%%%%%%%% Section 2. Hamiltonian structure: Rational $R$-matrix Poisson brackets %%%%%%%%%%%%%%%%

 \section{Hamiltonian structure: Rational $R$-matrix Poisson brackets}
   \subsection{Classical $R$-matrix theory}
   The results summarized in this section are basic ingredients of the classical $R$-matrix approach to integrable  isospectral Hamiltonian
   systems, realized on {the duals of} loop algebras \cite{Sem, AHH}.
 The rational $R$-matrix Poisson brackets on the phase space (also known as ``linear Leningrad brackets'') are defined  by
\be
\{L_{ab}(z), L_{cd}(w) \}= \frac{1}{z-w} \Big( (L_{ad}(z) -L_{ad}(w) )\delta_{cb}  -  (L_{cb}(z) -L_{cb}(w) )\delta_{ad}  \Big).
\ee
\noindent { Classical $R$-matrix theory \cite{Sem} then implies that:}
\begin{itemize}
\item
The elements of the ring $\II^{\Ad^*}(L^*\grgl(r))$ of
$\Ad^*$ invariant functions of $L(z)$ (i.e., the ring of spectral invariants), generated by the coefficients of the
 {\it characteristic polynomial} defining the {\it (planar) spectral curve}
\be
\det(L(z) - \lambda \Ib)=0,
\ee
 all {\it Poisson commute}
\be
\{f, g\} =0, \quad \forall \ f, g\in \II^{\Ad^*}(L^*\grgl(r)).
\ee
\item
The Hamiltonian vector field $\Xb_H$  generated by any element $H\in \II^{\Ad^*}(L^*\grgl(r))$ is given by a commutator 
\bea
\Xb_H(X)&\& =\{X, H \} = [R_s(d H), X], 
 \label{Hamvec_commut}\cr
 \forall \ H &\&\in \II^{\Ad^*}(L^*\grgl(r)), \ X \in L\grgl(r),
\eea
where $X \in L\grgl(r)$ is viewed as a {linear functional} on $L^*\grgl(r)$ under the 
 {trace-residue} pairing and $R_s$ is the endomorphism of $L\grgl(r)$ defined by
\be
R_s(Y_+ + Y_-) = s Y_+  + (s - 1) Y_-,    \quad Y \in L\grgl(r)
\ee
for any $s \in  \Cbb$. In particular,  
\be
R_1(Y_+ + Y_-) = Y_+, \quad  \text{and} \quad R_{0}(Y_+ + Y_-)= - Y_-.
\ee
\end{itemize}

%%%%%%%%  Section 3.  Hamiltonian structure of isomonodromic deformations %%%%

  \section{Hamiltonian structure of rational isomonodromic deformation equations}
  The results described in this section and the two subsequent ones are derived in detail in ref.~\cite{BHH}. 
They  generalize, to  all rational nonresonant systems of any rank, those first derived in \cite{H1} 
for systems with a finite number of Fuchsian singularities at finite points, plus one of Poincar\'e rank $1$  at $\infty$.
    \subsection{Birkhoff invariants as spectral invariants and Casimirs}
 \begin{theorem}[{\it Birkhoff invariants as Casimirs of the rational $R$-matrix structure and spectral invariants} \cite{BHH} ]
 
The matrix $ \frac {d  T^\nu(\zeta_\nu)}{d \zeta_\nu}$ equals the principal part of the local Laurent
series of the matrix $\Lambda^\nu(\zeta_\nu)=\diag(\lambda^\nu_1, \cdots, \lambda^\nu_r)$ of eigenvalues near $z=c_\nu$
\be
 \frac {d T^\nu}{d \zeta_\nu}(\zeta_\nu) =  \big(\Lambda^\nu(\zeta_\nu)\big)_{sing}
\ee
where, under the assumptions defining the Lax matrix $L(z)$ in eq.~(\ref{Ldef}),  the characteristic equation 
\be
\quad \det(L(z) - \lambda^\nu_a \Ib)=0   \quad \text{near } z=c_\nu , 
\ee
determines $r$ distinct solutions $\{\lambda^\nu_a\}_{a=1, \dots, r}$ as eigenvalues near to each singular point $z\in\{c_\nu, \infty\}_{\nu =1, \dots, N}$,
 which may be locally expressed as distinct Laurent series
\bea
\lambda^\nu_a(\zeta_\nu) &\&=  -\sum_{j=0}^{d_\nu}\frac{ t^\nu_{ja}}{\zeta_\nu^{j+1}} +   \OO(1),  \quad \nu=1, \dots, N,  
\cr
\lambda^\infty_a(\zeta_\infty) &\&=  \sum_{j=0}^{d_\infty}\frac{ t^\infty_{ja}}{\zeta_\infty^{j-1}} +   \OO(\zeta^2_\infty) 
\eea
in some neighbourhood of the singular points $z\in\{c_\nu, \infty\}_{\nu=1, \dots, N}$.
Therefore
\bea 
\qquad \quad  \quad t^\nu_{j  a} &\&=- \res_{z=c_\nu }\zeta_\nu^j \lambda^\nu_a(z) d z ,\cr
   \quad j =1, \dots, d_\nu,  &\& \quad \nu = 0, \dots, N, \infty, \quad a=1, \dots, r.
\eea
    The {Birkhoff invariants}  $\{t^\nu_{ja}\}_{\nu=1, \dots, N,\infty, \, j=1,\dots, d_\nu, \, a=1, \dots, r}$ ,
the {exponents of formal monodromy}  $\{t^\infty_{0a}\}_{j=1,\dots, d_\infty, a=1, \dots, r}$  at the finite poles and the {pole loci}
 $\{c_\nu\}_{\mu=1, \dots, N}$ are all {Casimir elements} of the Poisson structure. They are functionally independent, and generate the
center of the Poisson algebra; i.e., the ring of Casimir invariants. 
    \end{theorem}

    \subsection{Spectral invariant isomonodromic Hamiltonians}
 \begin{theorem}[{\it  Hamiltonians  as {\it dual spectral invariants}} \cite{BHH}]
In  punctured neighbourhoods of the singular points the eigenvalues have  distinct Laurent expansions of the form
 \bea
\lambda^\nu_a(\zeta_\nu) &\&= - \sum_{j=1}^{d_\nu}\frac{ t^\nu_{ja}}{\zeta_\nu^{j+1}}  -  \frac{ t^\nu_{0a}}{\zeta_\nu}
-  \sum_{j=1}^{d_\nu} jH_{t^\nu_{ja}} \zeta_\nu^{j-1}
+ \OO(\zeta_\nu^{d_\nu}),   \quad \nu=1, \dots, N, \\
\lambda^\infty_a(\zeta_\infty) &\&=  \sum_{j=1}^{d_\nu}\frac{ t^\infty_{ja}}{\zeta_\infty^{j-1}} 
+ t^\infty_{0a}\zeta_\infty  +\sum_{j=1}^{d_\infty} j H_{t^\infty_{ja}} \zeta_\infty^{j+1}
+ \OO(\zeta_\infty^{d_\infty+2}), 
\eea
where the Hamiltonians are 
 \bea
H_{t^\nu_{j a}} &\&:= -\frac 1{j} \res_{z=c_\nu} \frac{1}{ \zeta_\nu^{j}} \lambda_a(z)d z
={- \res_{z=c_\nu} \tr \left(( Y^\nu)^{-1} \frac {d  Y^\nu}{d z}  \frac {\pa T^\nu}{\pa t^\nu_{j a} }  \right)d z} , 
\label{Htnuja}\\
&\& \cr
 \nu&\&=1,\dots, N, \infty,  \quad j=1,\dots, d_\nu, \quad a= 1,\dots, r, \cr
 &\& \cr
 H_{c_\nu} &\&:=\frac 1 2 \res_{z=c_\nu}\tr \Big(L^2(z)\Big) d z 
 ={- \res_{z=c_\nu} \tr \left(( Y^\nu)^{-1} \frac {d Y^\nu }{d z}  \frac {\pa T^\nu }{\pa c_\nu } \right)d z} ,
 \label{Hcnu}
\eea
and the second equalities in eqs.~(\ref{Htnuja}),  (\ref{Hcnu}) hold when these spectral invariants are evaluated
on the solution manifolds of the isomonodromic deformation equations.
      \end{theorem}

\subsection{Isomonodromic $\tau$-function (Jimbo-Miwa-Ueno (1981))}

\noindent{\bf Differentials on the space of deformation parameters:}

\noindent Define:
 \bea
{\rm d}_\nu := d c_\nu \frac \pa{\pa c_\nu}+ \sum_{j=1}^{d_\nu} \sum_{a=1}^r 
d t  ^{\nu}_{ja} \frac \pa{\pa t^{\nu}_{ja}}, \quad
{\rm d}_\infty := \sum_{j=1}^{d_\infty} \sum_{a=1}^r  d t  ^{\infty}_{ja} \frac
\pa{\pa t^{\infty}_{j a}}\ .
\eea

\begin{theorem}[{\cite{JMU}}]
\noindent{The differential $1$-form:}
\be
\omega_{_{IM}} :=-\sum_{\nu=1}^{N, \infty} \res_{z=c_\nu}
 \Bigg(\tr \left((Y^{\nu}(\zeta_\nu))^{-1} \pa_z Y^{\nu}(\zeta_\nu) {\rm d}_\nu T^{\nu}(\zeta_\nu) \right)d \zeta_\nu\Bigg) 
\ee
is  {closed} when restricted to the solution manifold of the isomonodromic equations and hence {locally exact} \cite{JMU}. 
\end{theorem}

The {isomonodromic $\tau$-function} $\tau_{IM}$  is locally defined  \cite{JMU}, up to a parameter independent normalization, by 
\be
 \omega_{_{IM}} := {\rm d} \ln \tau_{_{IM}} 
=\sum_{\nu =1}^N H_\nu dc_\nu + \sum_{\nu=1}^{N, \infty}\sum_{j=1}^{d_\nu} \sum_{a=1}^r   H_{t^\nu_{ja}}  d t  ^{\nu}_{ja}.
\ee
Globally, it is a {section of a line bundle} over the space of deformation parameters 
\bea
\Tb&\&:= \{t^\mu_{ja}, c_\nu\}, \ 
\mu=1, \dots, N, \infty,  \ j=1, \dots d_\mu, \  \nu=1, \dots, N,  \ a=1, \dots, r .
 \nonumber
\eea
\subsection{Hamiltonian vector fields and explicit derivatives }
\begin{theorem}[{\it Hamiltonian vector fields} \cite{BHH} ]
The Hamiltonian vector fields corresponding to the spectral invariant Hamiltonians  $(H_{t^\nu_{ja}},H_{c_\nu})$  are given by the commutators
\be
\Xb_{H_{t^\nu_{ja}}} L:=\Big[U^\nu_{ja}, L\Big],   \quad 
\Xb_{H_{c^\nu}} L:=\Big[V^\nu,  L\Big],  
\ee
where
\bea
R_0(dH_{t^\nu_{ja}}) &\&=  U^\nu_{j  a }(z;L) = -(d H_{t^\nu_{ja}})_- , \cr
&\& \cr
R_0({dH_{c_\nu}})&\& =  V^\nu (z;L) =  -(d H_{c_\nu})_- \quad  \text{for}  \ \nu=1, \dots, N,\cr
&\& \cr
 R_1(dH_{t^\infty_{ja}})&\& = U^\infty_{ja} = (d H_{t^\infty_{ja}})_+.
\eea
\end{theorem}
\begin{definition} The {\it explicit derivatives} with respect to the deformation parameters 
are defined by the {\it Isomonodromic Conditions}:
\be
\nabla_{t^{\nu}_{ja}}L(z):=  \frac {d}{d z} U^\nu_{j  a }(z;L) ,\qquad
\nabla_{c_\nu}L(z):=  \frac {d}{d z} V^\nu (z;L)  .
\ee
\end{definition}
\noindent Adding these to the  Hamiltonian vector fields
\be
\Xb_{H_{t^\nu_{ja}}}+ \nabla_{t^\nu_{ja}}\ \text{and } \Xb_{H_{c_\nu}} + \nabla_{c_\nu}
\ee
 gives the {\it zero-curvature equations}
\bea
\frac{\partial L(z)}{\partial t^{\nu}_{ja}}&\&= \Big[U^\nu_{ja}, L\Big] +\frac {d  U^\nu_{j  a }(z)}{d z} \cr
\frac{\partial L(z)}{\partial c_\nu}
&\&= \Big[V^\nu, L\Big] +\frac {d V^\nu(z}{d z} .
\eea

\smallskip

\noindent These are the consistency conditions for the JMU equations \ref{JMUeq},
guaranteeing the invariance of the {generalized monodromy} (including the values of the {Stokes matrices})
under changes in the deformation parameters $\{t^\nu_{ja}, c_\nu\}$. 

The question is: in what sense are
\be
\nabla_{t^{\nu}_{ja}} = \frac{\partial^0}{\partial^0 t^\nu_{ja}}, \quad \nabla_{c_\nu} = \frac{\partial^0}{\partial^0 c_\nu}
\ee
``{explicit derivatives}'', defining a ``{trivial flat connection}''? This is justified in the next section.

%%%%%%%%% Section 5. Consistency conditions for explicit derivatives %%%%%

\section{Consistency conditions for explicit derivatives}
 
   \subsection{Explicit derivatives: consistency conditions, Poisson invariance }
   \begin{theorem}[{\it Consistency conditions for explicit derivatives} \cite{BHH}]
For all  $\mu, \nu=1,\dots, N$, and $\nu=\infty$,  the {explicit derivative} vector fields $\{\nabla_{c_\mu}, \nabla_{t^\nu_{ja}}\}_{ j=1,\dots d_\nu, a=1,\dots,r}$ 
{\it commute amongst themselves}, 
\be
[\nabla_t, \nabla_s] = 0, \quad  \forall \ t,s \in \Tb,
\ee
generating a (locally) {\it free abelian group action} that is transversal to the symplectic foliation, with
\be
\nabla_t(s) =0,  \quad  \forall \ t,s \in \Tb.
\ee
\end{theorem} 

\begin{theorem}[{\it Invariance of Poisson brackets under $\nabla_t $'s} \cite{BHH}]
Let $t$ denote any of the isomonodromic deformation parameters $t\in \Tb$ and  $\nabla_t$ be the corresponding  
{explicit derivative} vector field. Then
\be
\nabla_t\{f,g\} = \{\nabla_t f, g\}  + \{f, \nabla_t g\}.
\ee
In particular, if $f,g$ are in the joint kernel of all the $\nabla_t$'s,  their Poisson bracket $\{f, g\}$ is also.
\end{theorem}

\subsection{Transversality: Poisson quotient by abelian group action}

   Let 
\be
\TT:= \Span\{\nabla_t, \ t \in \Tb\}
\ee
 define the \noindent {\it transversal distribution.}
 \begin{theorem}[{\it Poisson quotient by abelian group action} \cite{BHH}]
 $\TT$ is an integrable distribution of constant, maximal rank
\be
N+ r \sum_{\nu=1}^N  d_\nu  + r d_\infty, 
\ee
 transversal to the symplectic foliation and the  canonical projection 
 $\pi:{\LL_{r, \db}} \to \WW := {\LL_{r, \db}}/ \TT$ is Poisson.
\end{theorem}

    \section{Examples. (See  refs.~\cite{BHH, H1, HB} for further details.)}
\subsection{Example 1. Schlesinger equations (Fuchsian).  Only first order poles in the Lax matrix (see refs.~\cite{H1, HB}).}
The covariant derivative equation is:
\be
\frac{ \partial \Psi(z)}{\partial z}=  L^{\text{Sch}}(z)\Psi(z),  
\ee
where
\be
L^{Sch}(z):= \sum_{\nu =1}^N \frac{L^{\nu}} {z- c_\nu},
\ee
and the deformation equations are
 \be
 \frac{\partial \Psi}{\partial c_\mu} = - \frac{L^\nu}{z-c_\nu}\Psi.
 \label{Schles_def_eqs}
 \ee
The {\it Schlesinger equations}
 \bea
\frac{\partial L^{\mu}}{\partial c_\nu} &\&= \frac{[L^{\mu}, L^\nu]}{c_\mu - c_\nu}, \quad  \forall \   \nu \neq  \mu, \cr
\frac{\partial L^{\mu}}{\partial c_\mu}  &\&= - \sum_{\nu=1, \ \mu \neq \nu}^N\frac{[L^{\mu}, L^\nu]}{c_\mu - c_\nu},
\label{Schles}
\eea
are the (zero-curvature) compatibilty conditions.
The Hamiltonians are
\be
H_\nu := \frac{1}{2}\res_{z= c_\nu} \tr \left(L^{Sch}\right)^2d z, 
\ee
and the $\tau$-function is determined from
\be
 d\ln(\tau^{Sch})= \sum_{\nu=1}^N H_\nu dc_\nu.
 \ee
 The infinitesimal isomonodromic deformation matrices are 
 \be
 R_0(d H_\nu) = - (d H_\nu)_- = - \frac{L^\nu}{z-c_\nu},
 \label{Schles_inf_isomon_mat}
 \ee
 and these  satisfy the {\it isomonodromic conditions} given by
 \be
\ \frac{\partial L^{\text{Sch}}}{\partial c_\nu}   =\frac{ \partial  \left(\frac{-L^\nu}{z-c_\nu}\right)}{\partial  z}  
= \frac{L^\nu}{(z-c_\nu)^2}.
\label{Schles_isomon_cond}
\ee

\subsection{Example 2 . Fuchsian,  plus double pole at $z=\infty$ (see  refs.~\cite{BHH, H1}).}

The covariant derivative equation is
 \be
 \frac{ \partial \Psi(z)}{\partial z} =  L^{\text{B}}(z)\Psi(z),  \quad \Psi(z)\in {\grGl}(r),
 \label{LB_eq}
 \ee
 where
 \be
L^B(z):= B + L^{Sch}(z), \quad B:= \diag(t^\infty_1, \dots, t^\infty_r), 
\ee
so there is a double pole in the $1$-form $L(z)dz$ at $z=\infty$. 
We again have the first order poles  at finite points $z=\{c_\nu\}_{\nu=1, \dots, N}$ and corresponding 
Schlesinger-like Hamiltonians
\be
H_\nu := \frac{1}{2}\res_{z= c_\nu} \tr \left(L^B\right)^2d z,
\ee
which generate the same set of deformation equations (\ref{Schles_def_eqs}) (with $L^{Sch}$ replaced by $L^B$)
and satisfy the same isomonodromic conditions (\ref{Schles_isomon_cond}).
In addition, the spectral curve and invariants at $z=\infty$ give the further Birkhoff invariants and spectral invariant
Hamiltonians $\{K_a\}_{a=1, \dots, r}$ defined by  
\bea
 \det(L^B(z) - \lambda^\infty_a \Ib)&\&=0, \cr
 &\&\cr
 t^\infty_{a}  = \res_{z=\infty } z^{-1} \lambda^\infty_a(z) d z , \quad
&\& K_a :=  \res_{z=\infty }z \lambda^\infty_a(z)d z,
\eea
which  all Poisson commute, together with the  $\{H_\nu\}_{\nu=1, \dots, N}$.
The  $K_a$'s satisfy the {\it isomonodromic conditions}
 \bea
\frac{\partial (dK_a)_+}{\partial z} = \frac{\partial^0 L^B}{\partial^0 t^\infty_a}  = E_a, \quad a=1, \dots, r,
\eea
where
\be
 (dK_a)_+ = \left(z E_a + \sum_{\nu=1}^N \sum_{b=1 \atop b\neq a}^r \frac {E_a L^\nu E_b + E_b L^\nu E_a}{t^\infty_a - t^\infty_b}\right),
\ee
whose compatibility with (\ref{LB_eq}) provide the zero curvature conditions that render the corresponding 
deformation equations 
\be
 \frac{\partial \Psi}{\partial {t^\infty_a}} =  (dK_a)_+ \Psi  \quad a=1, \dots, r
 \label{LB_def_eq}
 \ee
isomonodromic \cite{H1, JMU}.

\subsection{Example 3.  Hamiltonian structure of Painlev\'e  $P_{II}$ equation: $N=0$, \ $r=2$, \ $d_\infty =3$ (see refs.~\cite{BHH, HB}).}
The $P_{II}$ equation is:
\be
\frac{d^2 u}{dt^2} = 2 u^3 + t u + \alpha, \quad (\alpha = \text{const.})
\label{PII}
\ee
The linear system is:
\be
{\partial \Psi(z)\over \partial z} = L^{P_{II}}(z)\Psi(z)  , \quad
{\partial \Psi(z) \over \partial t} = U(z)\Psi(z),  
\ee
where
\bea
L^{P_{II}}(z) &\&:= z^2  \begin{pmatrix} 1 & 0 \cr 0 & -1 \end{pmatrix}
 +z \begin{pmatrix} 0 & - 2y_1 \cr x_2 & 0 \end{pmatrix}
+ \begin{pmatrix} x_2 y_1 + {t\over 2} & - 2 y_2 \cr x_1 & -x_2y_1- {t\over 2} \end{pmatrix}
\cr
&\& \cr
U(z)&\&:= {z\over 2}  \begin{pmatrix} 1 & 0 \cr 0 & -1 \end{pmatrix} 
 + {1\over 2}  \begin{pmatrix} 0 & -2y_1 \cr x_2 &  0 \end{pmatrix}, 
 \eea
 and
 \be
  t= \frac{1}{2}  \res_{z=0} z^{-3}\tr\left(((L^{P_{II}})^{P_{II}})^2(z)\right)d z  = 2t^\infty_{11},
  \ee
is the Birkhoff  Casimir invariant at $z=\infty$. 

The spectral invariants are defined by
\bea
\lambda &\&= \pm \sqrt {-\det(L)} = \pm\left (z^2 + \frac t 2 - \frac {x_1y_1 + x_2 y_2}{z} + \frac {H_{II}}{z^2} + \dots\right), \cr
&\& \cr
H_{II} &\&= \frac{1}{4}  \res_{z=0} z^{-1}\tr(L^2(z)) -\frac{t^2}{8} =\frac{1}{ 2} \left( x_2^2y_1^2 + t x_2 y_1 - 2 x_1 y_2\right) ,
\eea
and the {\it isomonodromic condition} is:
\be
\frac{\partial^0 L^{P_{II}}}{\partial^0 t}=\nabla_t(L^{P_{II}}) = {1\over 2} \begin{pmatrix} 1 & 0 \\ 0 & -1 \end{pmatrix} = \frac{\partial U}{\partial z}.
\ee

  Choosing {new canonical coordinates}:
\bea
u &\&:=\frac{x_1}{ x_2},\quad v:= x_2 y_1,   \quad w:= \ln x_2,  \quad a:=x_1y_1+ x_2y_2,\cr
\theta&\&= y_1dx_1 + y_2 dx_2 =v d u+ a d w, 
\eea
the Hamiltonian becomes 
 \be
H _{II}= \frac{1}{2 } v^2 + {1\over2} (t +2 u^2) v- a u . 
\ee
 The $\tau$-function is determined from 
 \be
 \quad d\ln (\tau) = H_{II}dt, 
 \ee
 and the autonomous spectral invariant is : 
\be
 a  = -\frac{1}{4}\res_{z=0} z^{-2} \tr(L(z))^2. 
\ee
$w$ is an ignorable coordinate, and Hamilton's equations become:  
\be
 \frac{d u}{d t }= v + u^2+{t\over 2}, \quad {d v \over d t  } = -2u v +a, 
 \ee
 which are equivalent to (\ref{PII}) with $ \alpha := a - 1/2$.

 \subsection{Example 4.  Higher order elements of the $P_{II}$ hierarchy: $N=0$, $r=2$, $d_\infty =4$ (see ref.~\cite{BHH}).  }
 
 The linear system is:
 \bea
 {\partial \Psi(z)\over \partial z} &\& = L^{P_{II,2}}(z)\Psi(z)  , \cr
 &\&\cr
{\partial \Psi(z) \over \partial t_1} &\&= U_1(z)\Psi(z) ,\quad
 {\partial \Psi(z) \over \partial t_2} = U_2(z)\Psi(z) 
 \eea
 where
 \bea
L^{P_{II, 2}}(z) &\&:=\left({z}^{3}+ \left(t_2-x_{1}\,y_{2} \right) z-x_{1}\,y_{3}-x_{3}\,y_{2}+t_{1} \right) \sigma_3\cr
&\&- \sqrt {2}\left( x_{1}\left( {z}^{2}+\frac{t_{2}} 2 \right) 
+x_{3}\,z+x_{2}  -\frac 1 4 y_{2}\,{x_{1}}^{2}\right) \sigma_+ \cr
&\& -\sqrt {2} \left( y_{2}\, \left( {z}^{2}+\frac{t_{2}}2 \right) +y_{3}\,z+y_{1}  -\frac 1 4x_{1}\,{y_{2}}^{2} \right) \sigma_-
\noindent
\eea
and the deformation matrices are:
\be
U_1 := \left[ \begin {array}{cc} z&-\sqrt {2}x_{1}\\ \noalign{\medskip}-
\sqrt {2}y_{2}&-z\end {array} \right],\quad
 U_2 := \frac 12 \left[ \begin {array}{cc} -x_{1}y_{2}+{z}^{2}&-
\sqrt {2} \left( x_{1}z+x_{3} \right) \\ \noalign{\medskip}-
\sqrt {2} \left( y_{2}z+y_{3} \right) &x_{1}y_{2}-{z}^{2}\end {array} \right] .
\ee

\noindent The spectral invariants are:
\be
\lambda = z^3 + t_2z  + t_1 + \frac {a}{z} + \frac {H_1}{2z^2} + \frac {H_2}{2z^3} + \mathcal O(z^{-4}) ,
\ee
and the exponent of formal monodromy at $z=\infty$ is
\be
t^\infty_0 := -\res_{z=\infty} \sqrt{-\det L(z)} d z= a := x_1y_1 + x_2 y_2 + x_3y_3.
\ee
The {\it isomonodromic conditions} are:
\bea
\frac{\partial^0 L^{P_{II,2}}}{\partial^0 t_1}&\&=\nabla_{t_1}(L^{P_{II,2}}) =  \begin{pmatrix} 1 & 0 \\ 0 & -1 \end{pmatrix} = \frac{\partial U_1}{\partial z}, \cr
\frac{\partial^0 L^{P_{II,2}}}{\partial^0 t_2}&\&=\nabla_{t_2}(L^{P_{II,2}}) = \begin{pmatrix} z & -\frac{x_1}{\sqrt{2} }\\ =- \frac{y_1}{\sqrt{2}} & -z \end{pmatrix} = \frac{\partial U_2}{\partial z}, 
\eea

Making a canonical change of coordinates,
\bea
x_1 &\& := u_1 {\rm e}^{w},\quad
x_2 := u_2 {\rm e}^{w},\quad 
x_3 := {\rm e}^{w}, \cr
y_1 &\& := v_1 {\rm e}^{-w},\ \ 
y_2 := v_2 {\rm e}^{-w},\ \ 
y_3 := (a - u_1 v_1 - u_2 v_2) {\rm e}^{-w},
\eea
 the canonical $1$-form becomes
 \be
\theta= \sum_{i=1}^3 y_i dx_i = v_1du_1 + v_2 du_2 + a dw,
\ee
and the reduced Hamiltonians are
\bea
H_1=   &\& \left( \frac3 2v_{2}{u_{1}}^{2}-t_{2}u_{1}+2u_{2} \right) a
-2 t_{1}u_{1}v_{2}
+ \left( {u_{1}}^{2}v_{1} +u_{1}u_{2}v_{2}-v_{2} \right) t_{2} \cr
&\&-\frac 32{u_{1}}^{3}v_{1}v_{2}
-\frac 32{u_{1}}^{2}u_{2}{v_{2}}^{2}
-2u_{1}u_{2}v_{1}
+\frac 32u_{1}{v_{2}}^{2}
-2{u_{2}}^{2}v_{2}+2v_{1}
\\
H_2  = &\& \frac 1 2{a}^{2}{u_{1}}^{2}
+ \left( -u_{1}t_{1}-t_{2}-u_{1} \left( {u_{1}}^{2}v_{1}+u_{1}u_{2}v_{2}-v_{2} \right)  \right) a
+ \left( {u_{1}}^{2}v_{1}+u_{1}u_{2}v_{2}-v_{2} \right) t_{1}
+\cr
&\&+\frac  14 {t_{2}}^{2}u_{1}v_{2}
+ \left( -\frac 14{v_{2}}^{2}{u_{1}}^{2}+\frac 1 2u_{1}v_{1}+\frac 1 2u_{2}v_{2} \right) t_{2}
+\frac 12{u_{1}}^{4}{v_{1}}^{2}
+{u_{1}}^{3}u_{2}v_{1}v_{2}
+\frac 1{16}{v_{2}}^{3}{u_{1}}^{3}
+\cr
&\&+\frac 1 2{u_{1}}^{2}{u_{2}}^{2}{v_{2}}^{2}
-\frac 54{u_{1}}^{2}v_{1}v_{2}
-\frac 54 u_{1}u_{2}{v_{2}}^{2}
+\frac 1 2{v_{2}}^{2}+u_{2}v_{1},
\eea
where $w$ is again an ignorable variable. The canonically conjugate variable $a$, which is the exponent of formal 
monodromy at $\infty$, is a conserved quantity.
The isomonodromic deformation equations are then Hamiltonian's equations for the time-dependent Hamiltonians  $H_1$ and $H_2$.

%%%%%%% Section 6. Further  developments: Darboux coordinates %%%%%%

\section{ Further  developments. Darboux coordinates} 

To express all higher isomonodromic deformation equations explicitly in Hamiltonian form we would need,
in addition to the Casimir invariant coordinate functions  $\{t^\nu_{ja}, c_\nu\}$, a set of Darboux (canonical) coordinates
$\{u_\alpha, v_\alpha\}_{\alpha= 1, \dots.K }$ on the {symplectic leaves} that are {invariant under the integrable distribution $\TT$}
corresponding to the trivial (flat) connection $\nabla$ defining the {explicit} derivatives of $L$
\bea
\nabla_t u_\alpha &\&=0, \quad \nabla_t v_\alpha = 0, \quad \forall \  \alpha=1, \dots , K\cr
2K &\&:=  r(r-1)\left(d_\infty +\sum_{\nu=1}^N d_\nu +N -1\right).
\eea

Progress in this direction was made by Marchal, Orantin and  Alameddine \cite{MOA, MA} for rank $r=2$, using  the {\it spectral  Darboux coordinates}
of \cite{AHH}, a {different trivialization} of the bundle, and different choices of Hamiltonians. 
To relate the two, a multi-time dependent {canonical transformation} is required.  

Other work on the Hamiltonian structure of rational isomonodromic deformations systems 
includes that of Yamakawa \cite{Yam}, who developed the general nonresonant rational isomonodromic deformation system 
along lines similar to that considered here, Mazzocco and Mo \cite{MaMo}, who derived the  Hamiltonian structure 
of the $P_{II}$ hierarchy using the twisted  loop algebra $L\grsl^{(1)}(2)$ to define the phase space,
 and Gaiur {\it et al.}~\cite{GMR}, who obtained higher order singularities in isomonodromic deformation systems 
 through coalescence of poles.

%%%%%%%%%%%%%%%%%%%%% Acknowledgements %%%%%%%%%%%%%%%%%
 \bigskip
\noindent 
\small{ {\it Acknowledgements.} The results presented here are  based on joint work with M.~Bertola and J.~Hurtubise \cite{BHH},
 which extended earlier results of ref.~\cite{H1}  to general nonresonant rational isomonodromic deformation systems.  See also
\cite{H2} for an earlier introductory presentation announcing some of these results, and additional input by G. Pusztai on perturbative aspects
of the theory. 
This work was partially supported by the Natural Sciences and Engineering Research Council of Canada (NSERC). }

 %%%%%%%%%%%%%%%% Bibliography %%%%%%%%%%%%%%%%
 
 \newcommand{\arxiv}[1]{\href{http://arxiv.org/abs/#1}{arXiv:{#1}}}


\begin{thebibliography}{99}

\bibitem{AHH} M.R.~Adams, J.~Harnad and J.~Hurtubise, ``Darboux Coordinates and Liouville-Arnold 
integration in Loop Algebras'',  {\it Commun.~Math.~Phys.} {\bf  155}, 385-413 (1993).

\bibitem{BJL} W.~Balser, W.B.~Jurkatz and D.A.~Lutz, ``Birkhoff Invariants and Stokes' Multipliers for Meromorphic Linear Differential Equations'', 
{\it J.~Math.~Anal.~Appl.} {\bf 71}, 48-94, (1979).

  \bibitem{BHH} M.~Bertola, J.~Harnad and J.~Hurtubise, ``Hamiltonian structure of rational isomonodromic deformation systems'',
 {\it J.~Math.~Phys.} {\bf 64}, 083502 (2023).

\bibitem{Birk} G.D.~Birkhoff, ``The Generalized Riemann Problem for Linear Differential Equations and the Allied Problems
for Linear Difference and q-Difference Equations'', {\it Proc.~Amer.~Acad.~Arts and Sciences} {\bf 49} (9), 521-568 (1913).

\bibitem{Bo1} P.~Boalch, ``Symplectic Manifolds and Isomonodromic Deformations'',
{\it Adv.~Math.} {\bf 163}, 137-205 (2001).

\bibitem{Bo2} P.~Boalch, ``Quasi-Hamiltonian geometry of meromorphic connections'',
{\it Duke Math.~J.} {\bf 139},  369-405 (2007).

\bibitem{FN1} H.~Flaschka and A.C.~Newell,  ``Monodromy and spectrum-preserving deformations I.''
{\it Commun.~Math.~Phys.}  {\bf 76},  65-116 (1980).

\bibitem{FN2}  H.~Flaschka and A.C.~Newell,  “The inverse monodromy transform is a canonical transformation” in: Nonlinear
Problems: Present and Future (Los Alamos,NM; 1981); North Holland Math. Stud. {\bf 61}, 65–89,
North Holland;, Amsterdam  (1982).

 \bibitem{Fu} R. Fuchs, ``\"Uber lineare homogene Differentialgleichungen zweiter Ordnung mit drei im Endlichen gelegene 
 wesentlich singul\"ere Stellen'', {\it Math.~Ann.} {\bf  63},   301-321 (1907).  
 
   \bibitem{GMR} I.~Gaiur, M.~Mazzocco and V.~Rubtsov,  ``Isomonodromic Deformations: Confluence, Reduction and Quantisation'', 
{\it Commun.~Math.~Phys.}  {\bf  400}, 1385-1461 (2023). 

 \bibitem{Gam} B.~Gambier. ``Sur les \'equations diff\'erentielles du second ordre et du premier degr\'e  dont l’int\'egrale g\'en\'erale est \`a points critiques fixes'', 
{\it Acta Math.}, {\bf  33}, 1–55, (1910).

\bibitem{Gar} R.~Garnier, ``Sur les \'equations diff\'erentielles du troisi\`eme ordre dont l’int\'egrale g\'en\'erale est uniforme et sur une classe d’\'equations 
nouvelles d’ordre sup\'erieur dont l’intégrale g\'en\'erale a ses points critiques fixes'', {\it Ann.~Ecol.~Norm.~Sup.}  {\bf 29}, 1-126 (1912) 

 \bibitem{H1} J.~Harnad, ``Dual Isomonodromic Deformations and Moment Maps into Loop Algebras'', 	
{\it Commun.~Math.~Phys.}  {\bf 166}, 337-365 (1994).

 \bibitem{H2} J. Harnad, ``Hamiltonian theory of the general rational isomonodromic deformation problem'',
presentation at Fields Institute workshop on integrable and near-integrable Hamiltonian systems,
May 17 -21, 2004, URL:  http://www.fields.utoronto.ca/audio/03-04/integrable/harnad/ .
    
  \bibitem{HB}   J.~Harnad and F.~Balogh, {\em Tau Functions and their Applications}, Monographs on Mathematical Physics series,  Cambridge University Press, Cambridge, UK  (2021).

\bibitem{JMU} M.~Jimbo, T. Miwa and K. Ueno,
 ``Monodromy preserving deformation of linear ordinary differential equations with rational coefficients.
 General theory and $\tau$-function'',  {\it Physica} {\bf 2D},  306-352 (1981).
 
 \bibitem{JM} M.~Jimbo and T.~Miwa, ``Monodromy preserving deformation of linear ordinary differential equations
with rational coefficients. II'',  {\it Physica} {\bf 2D},  407-448 (1981).

\bibitem{Malm} J.~Malmquist, ``Sur les \'equations diff\'erentielles du second ordre dont l’int\'egrale 
g\'en\'eral a ses points critiques fixes'',  {\it Ark.~Mat.~Astr.~Fys.} {\bf 17}, 1-89 (1922-23).

\bibitem{MOA} O.~Marchal, N.~Orantin and M.~Alameddine,``Hamiltonian representation of isomonodromic 
deformations of general rational connections on $\grgl(2,\Cbb)$'', arXiv:2212.04833.

\bibitem{MA}  O.~Marchal and M.~Alameddine, ``Isomonodromic and isospectral deformations of 
meromorphic connections: the $sl_2(\Cb)$ case'', arXiv:2306.07378.

\bibitem{MaMo} M.~Mazzocco and M.Y.~Mo, ``The Hamiltonian structure of the second Painlev\'e\ hierarchy'', {\it Nonlinearity} {\bf 20}, (2007), 2845--2882.

\bibitem{Ok} K.~Okamoto, ``Polynomial Hamiltonians associated with Painlev\'e equations. I.''
{\it  Proc.~Japan Acad.} {\bf 6}, 264-268 (1980).

\bibitem{Pa1} P.~Painlev\'e, ``M\'emoire sur les \'equations diff\'erentielles dont l’int\'egrale generale est uniforme'',
 {\it Bull.~Soc.~Math.~France} {\bf 28}, 201-261 (1900).

 \bibitem{Pa2} P.~Painlev\'e, ``Sur les \'equations diff\'erentielles du second ordre aux points critiques fixes. 
 {\it C. R. Acad. Sc. Paris} {\bf 143}, 1111-1117 (1906).

\bibitem{Pic} E. Picard, ``M\'emoire sur la th\'eorie des functions alg\'ebriques de deux variable'',
 {\it J. Liouville} {\bf 5}, 135-319 (1889). 
 
  \bibitem{Sch}  L.~Schlesinger,  ``\"Uber eine Klasse von Differentialsystemen beliebiger Ordnung mit festen kritischen Punkten''{,
\it J.~Reine u.~Angew.~Math.} {\bf 141}, 96-145 (1912).

\bibitem{Sem}  M.A.~Semenov-Tian-Shansky,  ``What is a Classical R-Matrix'',
{\it Funct.~Anal.~Appl.} {\bf 17}, 259-272 (1983)

\bibitem{Yam} D.~Yamakawa, ``Tau functions and Hamiltonians of isomonodromic
deformations'', {\it Josai Math.~Monog.} {\bf 10}, 139160 (2017).
 
 
 \bigskip
\noindent


\end{thebibliography}
\end{document}